\documentclass[conference]{IEEEtran}
\IEEEoverridecommandlockouts
\usepackage{cite}

\usepackage{amsmath,amssymb,amsfonts}
\usepackage{algorithmic}
\usepackage{graphicx}
\usepackage{textcomp}
\usepackage{xcolor}

\makeatletter
\IEEEtriggercmd{\reset@font\normalfont\fontsize{7.9pt}{8.40pt}\selectfont}
\makeatother
\IEEEtriggeratref{1}

\graphicspath{{../pdf/}{../jpeg/}{../eps/}}
\graphicspath{{Figures/}}

\def\BibTeX{{\rm B\kern-.05em{\sc i\kern-.025em b}\kern-.08em
  T\kern-.1667em\lower.7ex\hbox{E}\kern-.125emX}}
\begin{document}
\bstctlcite{IEEE_ICC:BSTcontrol}

\title{Experimental Demonstration of Staggered CAP Modulation for Low Bandwidth Red-Emitting Polymer-LED based Visible Light Communications\\
\thanks{This work was supported by the UK EPSRC grant EP/P006280/1: Multifunctional Polymer Light-Emitting Diodes with Visible Light Communications (MARVEL). EW acknowledges the Swedish Research Council and the Swedish Research Council Formas, the Wallenberg Foundation through the Wallenberg Academy Fellows program for financial support. FC is a recipient of a Royal Society Wolfson Foundation Research Merit Award.}
}

\author{Paul Anthony Haigh\textsuperscript{1}, Alessandro Minotto\textsuperscript{2}, Andrew Burton\textsuperscript{3}, Zabih Ghassemlooy\textsuperscript{3},\\Petri Murto\textsuperscript{4,5}, Zewdneh Genene\textsuperscript{4,6}, Wendimagegn Mammo\textsuperscript{6}, Mats R. Andersson\textsuperscript{5},\\Ergang Wang\textsuperscript{4}, Franco Cacialli\textsuperscript{2} and Izzat Darwazeh\textsuperscript{1}

\\
\IEEEauthorblockA{
\textit{\textsuperscript{1}Communications and Information Systems Group, University College London, Gower Street, WC1E~6BT, United Kingdom}\\
\textit{\textsuperscript{2}2Department of Physics and Astronomy, University College London, Gower Street, London, WC1E 6BT, UK}\\
\textit{\textsuperscript{3}Optical Communications Research Group, Northumbria University, Newcastle-upon-Tyne, NE1 8ST, UK}\\
\textit{\textsuperscript{4}Department of Chemistry and Chemical Engineering/Applied Chemistry, Chalmers University of Technology, Sweden}\\
\textit{\textsuperscript{5}Flinders Institute for NanoScale Science \& Technology, Flinders University, Adelaide, SA 5042, Australia}\\
\textit{\textsuperscript{6}Department of Chemistry, Addis Ababa University, Addis Ababa, P.O. Box 33658, Ethiopia}\\
\{p.haigh;a.minotto;f.cacialli;i.darwazeh\}@ucl.ac.uk; \{andrew2.burton;z.ghassemlooy\}@northumrbia.ac.uk}
}

\maketitle

\begin{abstract}
In this paper we experimentally demonstrate, for the first time, staggered carrier-less amplitude and phase (sCAP) modulation for visible light communication systems based on polymer light-emitting diodes emitting at $\sim$639~nm. The key advantage offered by sCAP in comparison to conventional multi-band CAP is its full use of the available spectrum. In this work, we compare sCAP, which utilises four orthogonal filters to generate the signal, with a conventional 4-band multi-CAP system and on-off keying (OOK). We transmit each modulation format with equal energy and present a record un-coded transmission speed of $\sim$6 Mb/s. This represents gains of 25\% and 65\% over the achievable rate using 4-CAP and OOK, respectively.
\end{abstract}


\section{Introduction}
Organic polymer light-emitting diodes (PLEDs) are emerging as a popular choice for visible light communications (VLC) in recent years. This has been driven by their ability to be dissolved into solvents and patterned using wet methods such as inkjet printing \cite{degans}, spin-coating \cite{yimsiri} or spray coating \cite{gaynor}; all of which can be performed at extremely low costs. Moreover, since there is no epitaxial crystal growth, arbitrary photoactive areas can be produced that are restricted only by the limitations of the manufacturing equipment. A further advantage of PLEDs is that they can easily be fabricated for any wavelength in the visible range of the electromagnetic spectrum, due to an abundance of polymers with a 1--4~eV band-gap \cite{clayden}.

Due to these advantages, and others such as mechanical flexibility, PLEDs are being adopted as the transmitter in point-to-point VLC links \cite{clark}. In recent years, one of the key challenges, which has widely been investigated by the research community, has been in increasing transmission speeds. Reports have shown that the data traffic growth is approximately exponential and mobile data traffic alone is expected to increase to 49~EB/month by 2021. Almost 75\% of this data will be generated by smart devices \cite{cisco}, which is the one of key target market for VLC utilisation. The most common method for increasing data rates is a combination of analogue and digital equalisation with advanced modulation formats and wavelength division multiplexing (WDM) \cite{wang2}. Using inorganic devices in VLC systems, data rates up to $\sim$10~Gb/s have been reported \cite{wang2,chun,islim,wang}.

Comparatively, data rates using PLEDs are limited due to their inherently lower modulation bandwidth. The root cause of the lower modulation bandwidths is that organic semiconductors have a lower charge mobility than inorganics by several orders of magnitude; hole mobilities are typically in the range of 10\textsuperscript{-6}--10\textsuperscript{-2} cm\textsuperscript{2}/Vs, with lower mobilities for electrons. This, therefore, severely limits the switching speed of PLEDs due to slow charge extraction and hence electroluminescence extinction, thus restricting their applications in broadcasting technologies. However, high-end display technologies with organic LEDs are currently emerging, which offer advantages such as superior switching speeds and no requirement for colour filters since they are a light source over liquid crystal displays (LCDs). The general interest in using PLEDs as a display technology is supported by the fact that the expected market value of flexible organic displays in 2025 is predicted to reach more than \$58B \cite{das}. Therefore, it is envisaged that, PLEDs can be used as an embedded uplink in smart devices and other display technologies, or as a device-to-device communications link.

Typically in modern wireless access network technologies, data rates are lower for the uplink and hence, a reduced transmit power level is allocated \cite{boccardi}. In general, VLC systems have usually been contextualised as a unidirectional broadcasting technology to date, with alternative technologies such as radio frequency (RF) being used for the uplink \cite{shao,shao2}. Due to the relaxed requirement for high data rates, it is envisaged that individually addressable pixels embedded into PLED display technologies can effectively be utilised as the uplink in future VLC networks whilst displaying an image. There is still a need however, to maximise available transmission speeds in organic-based VLC systems. In \cite{chen}, a world record data rate of 51.6~Mb/s was reported using a combination of a single pixel and offset-quadrature amplitude modulation (O-QAM)-based orthogonal frequency division multiplexing (OFDM) and linear minimum mean square error decision feedback equalisation. In \cite{haigh3}, a wavelength-multiplexed transmission link using three PLEDs transmitting independent data and a multi-layer perceptron based equaliser with an aggregated data rate of 55~Mb/s was reported. In terms of un-equalised transmission speeds, 4.5~Mb/s was reported in \cite{chen}. Increasing the data rate without using equalisation is an attractive prospect for smart devices and displays modules. High performance equalisers such as the multi-layer perceptron are computationally complex and require a large digital signal processing area, which will negatively impact battery life \cite{gurram}. Advanced multi-level and multi-carrier modulation schemes are one of the most effective methods for improving spectral usage (i.e., increasing spectral efficiency). In \cite{wu}, carrier-less amplitude and phase (CAP) modulation was demonstrated to outperform other formats under the same energy and bandwidth constraints. LEDs are low-pass filters in nature with significant high frequency attenuation. CAP can be extended to a multi-band scheme, which further improves spectral efficiency by mitigating this effect \cite{haigh}. The reason for this is that each sub-band experiences less attenuation than a conventional 1-band CAP system, hence signal to noise ratio (SNR)-per-sub-band is improved, thus leading to higher data rates.

The problem with CAP modulation in general, however, is the use of a guard-band, which is associated with the roll-off factor of the pulse-shaping filters, in order to combat inter-symbol interference (ISI) at the cost of lowering the data rate. Recently in \cite{stepniak}, a new method of generating a CAP signal known as staggered CAP (sCAP) was proposed, which offers full spectral usage by including two additional pulse amplitude modulation (PAM) signals with the conventional CAP system. In sCAP orthogonality is maintained by introducing a $T/2$ offset between the conventional CAP filters following the concept of O-QAM \cite{zhao}, where $T$ is the symbol period, which leads to zero interference at the sampling instance. It was shown that sCAP outperforms 4-band CAP in terms of bit error rate (BER) performance. Hence, in this work we have adopted sCAP in a polymer-based VLC link in an effort to improve the achievable transmission speed. The current record with no equalisation is 4.45~Mb/s using OFDM and the 7\% forward error correction (FEC) code \cite{chen}, and 3~Mb/s for un-coded links \cite{haigh3}. In this work, we demonstrate that, by employing sCAP in VLC data rates can be increased by $\sim$30\% and $\sim$65\%, compared to OFDM and the benchmark OOK, respectively with equivalent energy.

The paper is organised as follows. Section II describes the principle of sCAP modulation and the experimental test setup. Results are presented and discussed in Section III and finally, conclusions are given in Section IV.
\begin{figure*}[th]
  \centering
  \includegraphics[width=0.825\textwidth]{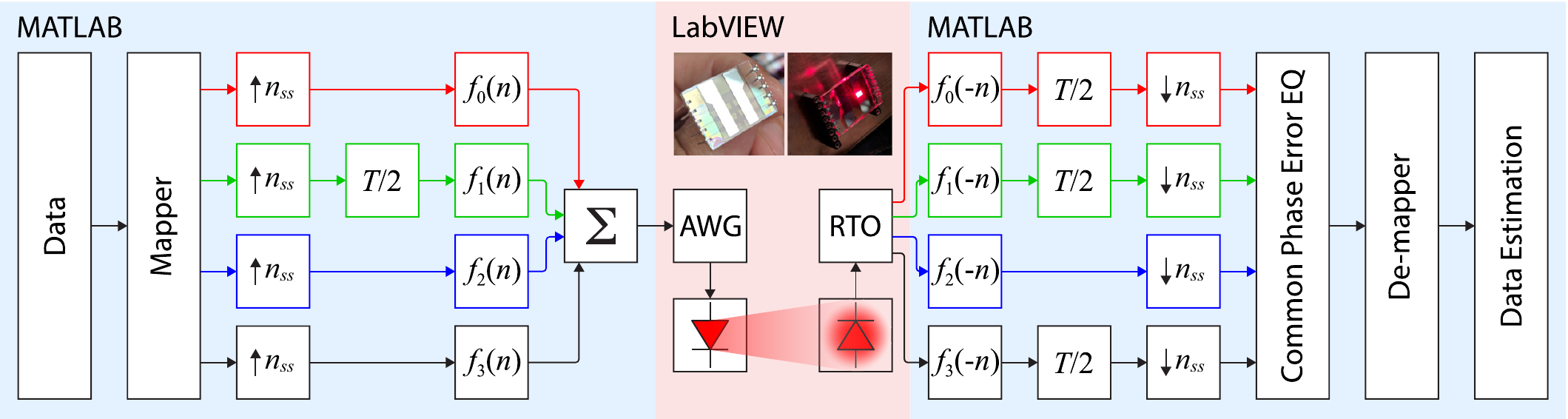}
  \caption{Simplified block diagram of the sCAP modulation format}
  \label{fig:figure1}
\end{figure*}

\section{Communication System under Test}
\begin{figure}[t]
  \centering
  \includegraphics[width=0.45\textwidth]{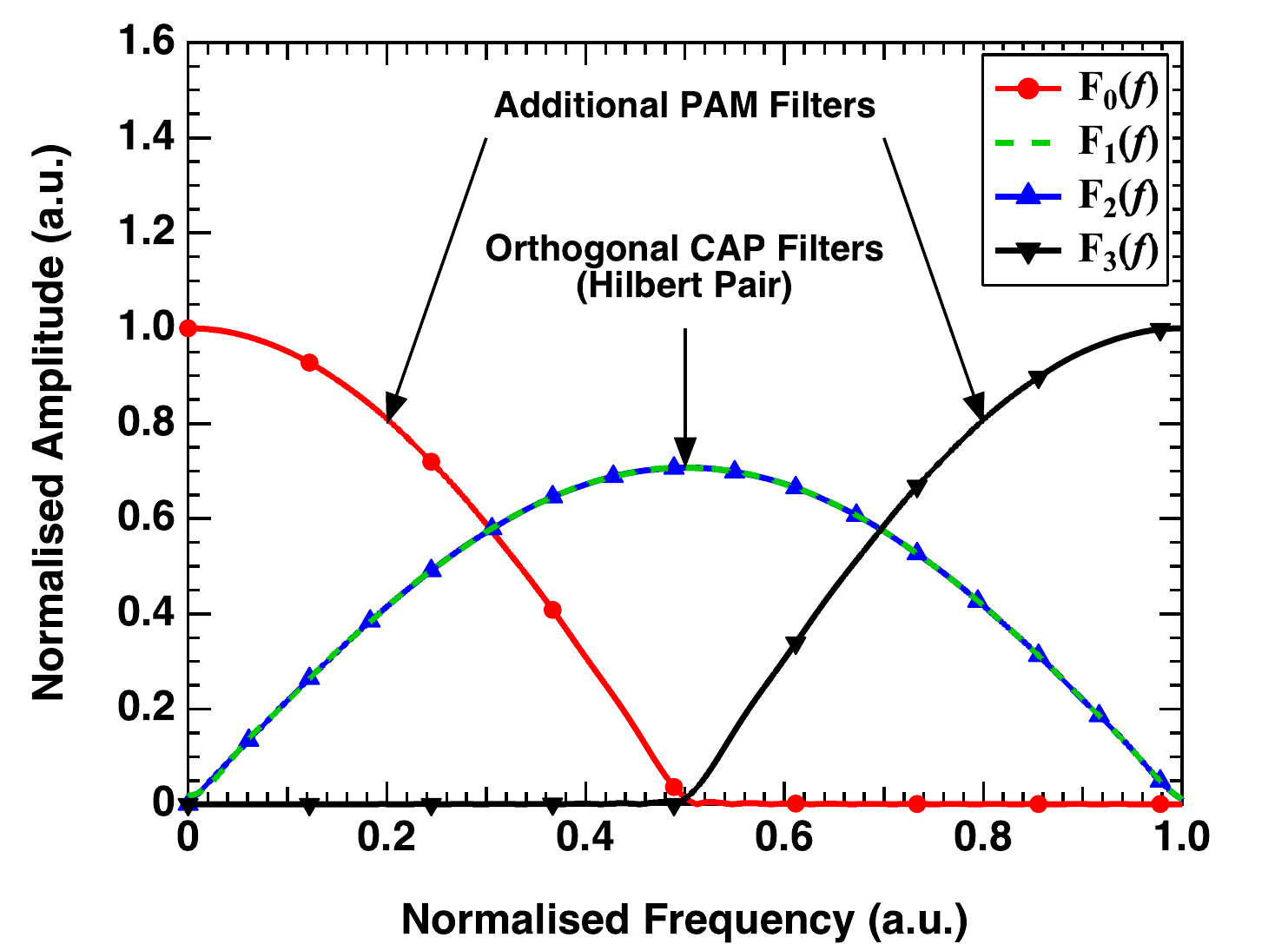}
  \caption{Ideal frequency response of the sCAP component signals, normalised to the Nyquist frequency}
  \label{fig:figure2}
\end{figure}

This section is composed of three subsections of (\emph{A}) the sCAP modulator, (\emph{B}) physical components used in the experiment and (\emph{C}) the sCAP demodulator.  The experimental system block diagram of the proposed sCAP PLED-VLC system is illustrated in Fig.~\ref{fig:figure1}. The sCAP system is designed using four orthogonal filters; $f_0(n)$, $f_1(n)$, $f_2(n)$ and $f_3(n)$, where $n$ is the current sample instance. The filters $f_1(n)$ and $f_2(n)$, are the equivalent to the Hilbert pair found in conventional CAP, i.e. they are the in-phase ($I$) and quadrature ($Q$) components which are orthogonal in time, occupy the same frequency space and are separated in phase by $\pi/2^c$. The key difference between sCAP and CAP is the addition of two PAM signals, which are shaped by $f_0(n)$ and $f_3(n)$ and placed either side of the CAP filters, as illustrated in the frequency domain in Fig.~\ref{fig:figure2}, that ensure full spectral usage. The sCAP system is compared with conventional 4-sub-band CAP and OOK transmission links. Block diagrams of these systems can be found in \cite{haigh,olmedo,haigh2}.
\subsection{Staggered-CAP Modulation}
A pseudorandom binary data sequence of length $2^{15}-1$ is generated in MATLAB and mapped onto the PAM alphabet prior to being up-sampled by zero-padding at a rate $n_{ss}~=~\chi\lceil2\left(1+\beta\right)\rceil$ where $\chi$ is an oversampling factor, and $\beta$ is the excess bandwidth factor. For sCAP, $\beta$ is set to unity, unlike generic CAP where $0<\beta<1$ \cite{haigh,olmedo}. In line with O-QAM, to maintain orthogonality between all filters, a $T/2$ offset is introduced between the $I$ and $Q$ components of $f_1(n)$ and $f_2(n)$. The reason for this, as reported in \cite{stepniak}, is that the cross-correlations for $f_0(n)$, $f_2(n)$ and $f_3(n)$ are zero at the sampling instance, while $f_1(n)$ has non-zero cross-correlation unless a $T/2$ offset is introduced.

\begin{figure}[t]
  \centering
  \includegraphics[width=0.45\textwidth]{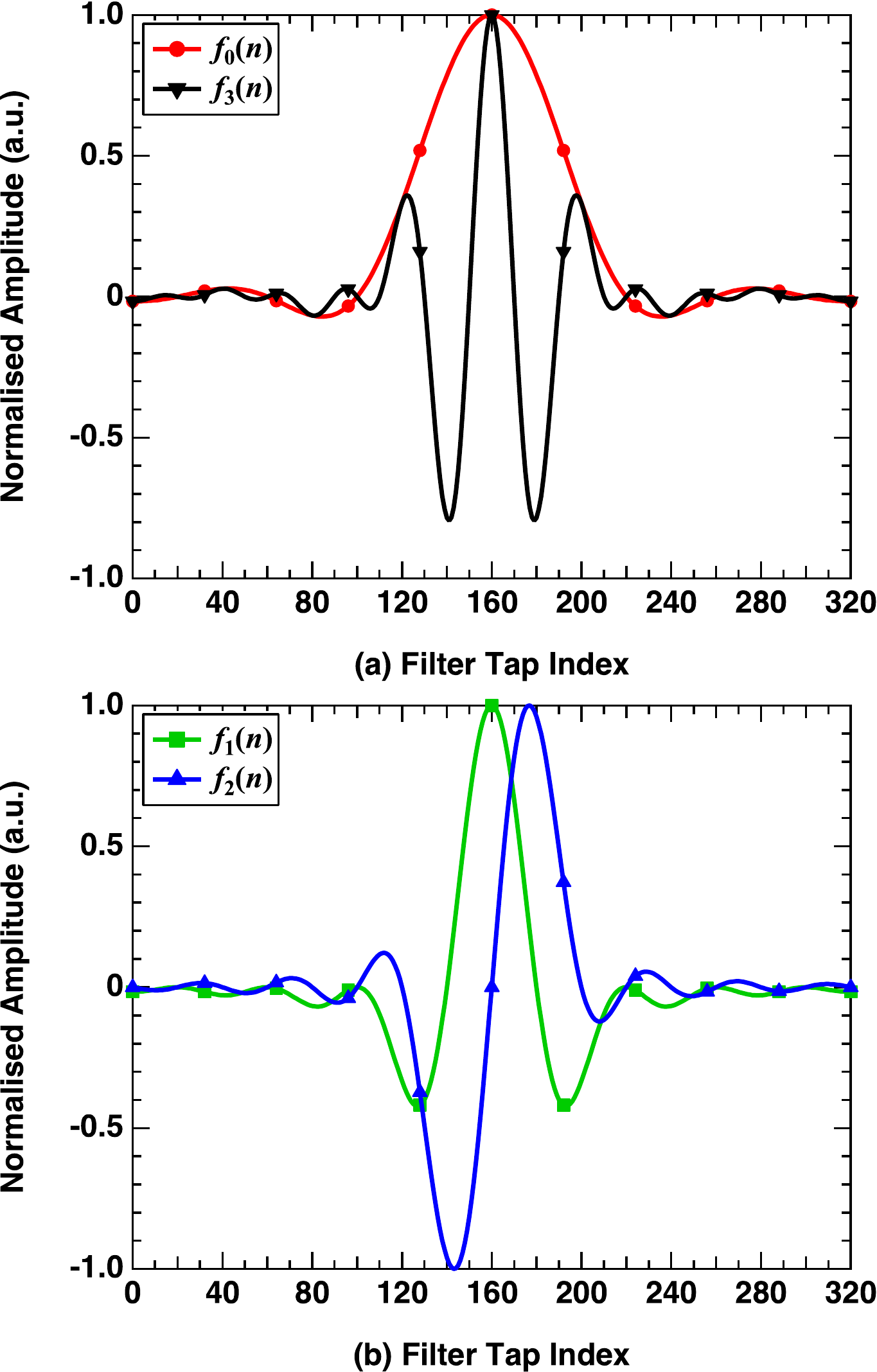}
  \caption{(a) The PAM and (b) CAP pulses that form the basis functions for sCAP}
  \label{fig:figure3}
\end{figure}

The first filter is a root-raised cosine (RRC), given as \cite{slim}:
\begin{equation}
  f_0(n)=\frac{4\beta\frac{n}{T}\cos{\left(\xi\left[1+\beta\right]\right)+\sin{\left(\xi\left[1-\beta\right]\right)}}}{\xi\left[1-\left(4\beta\frac{n}{T}\right)^2\right]}
  \label{eq1}
\end{equation}
where $\xi=\pi nT^{-1}$. L'H\^opital's rule must also be applied to remove the three singularities, following \cite{joost}. The remaining filters are based on $f_0(n)$ and are described as follows, starting with the central filters that form the Hilbert pair \cite{stepniak}:
\begin{equation}
  f_1(n)=\sqrt{2}f_0(n)\cos{\left(2\pi f_{c1}n\right)}
  \label{eq2}
\end{equation}
\begin{equation}
  f_2(n)=\sqrt{2}f_0(n)\sin{\left(2\pi f_{c2}n\right)}
  \label{eq3}
\end{equation}
where $f_{c1}$ and $f_{c2}$ are the carrier frequencies of the filters and $f_{c1}=f_{c2}=0.5B_{sub}$, where $B_{sub}$ is the bandwidth of the centre sub-bands. The final filter is the up-converted version of $f_0(n)$ as given by \cite{stepniak}:
\begin{equation}
  f_3(n)=f_0(n)\cos{\left(2\pi f_{c3}n\right)}
  \label{eq4}
\end{equation}
where $f_{c3}=2f_{c1}$. The unity-normalised filters are illustrated in Fig.~\ref{fig:figure3}, where Fig.~\ref{fig:figure3}(a) shows the impulse responses of $f_0(n)$ and $f_3(n)$, which are used for adding the additional PAM signals, whereas the centre filters $f_1(n)$ and $f_2(n)$, which form the Hilbert pair, are shown in Fig.~\ref{fig:figure3}(b). The combined four signals are loaded into the arbitrary waveform generator (AWG, Tektronix 70002A) using LabVIEW for intensity modulation of the PLED.

In addition, we generated (\emph{i}) the equivalent conventional 4-CAP using 4 pairs of orthogonal filters at the transmitter following the literature \cite{olmedo} with $\beta=0.1$ in order to maintain high spectral efficiency \cite{stepniak}; and (\emph{ii}) an OOK link \cite{haigh2} for comparison. Note that, for fair comparison we have adopted the same energy (i.e. SNR) for all modulation schemes.

\subsection{PLED under test}
The PLEDs used in this work were fabricated using glass substrates patterned with indium-tin oxide (ITO) as transparent anodes (OSSILA Ltd). Such substrates were cleaned with acetone and isopropanol in an ultrasonic bath and treated in an O\textsubscript{2} plasma chamber for 10 minutes to reduce the roughness of the ITO layer and increase the work function. To facilitate hole-injection, a 40~nm layer of poly(3,4-ethylenedioxythiophene)-poly(styrenesulfonate) (PEDOT:PSS) was deposited via spin-coating from a water dispersion (Heraeus Clevios AI 4083) in air and annealed at 150$^\circ$C for 10 minutes in a nitrogen-filled glovebox \cite{brown}. On top of the annealed PEDOT:PSS, a 100~nm thick active layer of poly[4,4,9,9-tetrakis(4-hexylphenyl)-4,9-dihydro-\emph{s}-indaceno[1,2-\emph{b}:5,6-\emph{b}']dithiophene2,7-diyl-\emph{alt}-5,5'-bis(2-octyldodecyl)-4\emph{H},4'\emph{H}-[1,1'-bithieno[3,4-c]pyrrole]- 4,4',6,6'(5\emph{H},5'\emph{H})-tetrone-3,3'-diyl] (PIDT-2TPD) with a number average molecular weight of 63.7~kDa and a polydispersity index of 2.3 was spin-coated from a 10~mgmL\textsuperscript{-1} toluene solution. Details about the synthesis of the polymer PIDT-2TPD have been reported in a previous work \cite{minotto}. Following this, a Ca/Al (30/200~nm) cathode was thermally evaporated on top in a high-vacuum chamber. Finally, the PLEDs were encapsulated by sandwiching a drop-cast epoxy glue (OSSILA Ltd) layer between the device and a protective glass slide (OSSILA Ltd). The glue was then cured by exposing the PLEDs under UV light for 15~minutes in the glovebox. The photoactive area of the PLED is 4.5~mm\textsuperscript{2}-per-pixel, and its cross-sectional schematic is shown in Fig.~\ref{fig:figure4}, with the molecular structures shown below. For photographs of the PLED see the inset in Fig.~\ref{fig:figure1}.
\begin{figure}[t]
  \centering
  \includegraphics[width=0.45\textwidth]{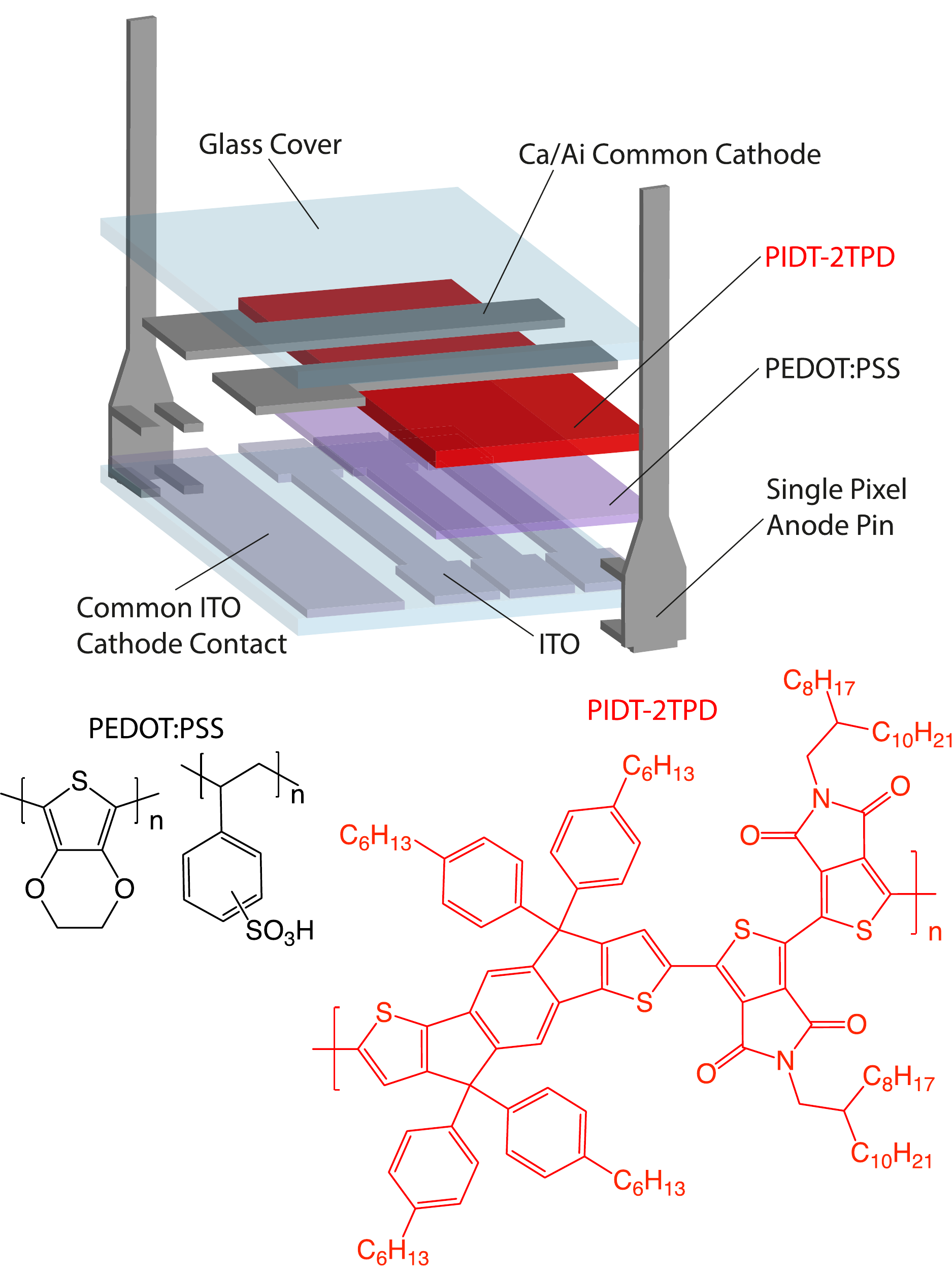}
  \caption{PLED materials layout and polymers used}
  \label{fig:figure4}
\end{figure}

\begin{figure}[t]
  \centering
  \includegraphics[width=0.45\textwidth]{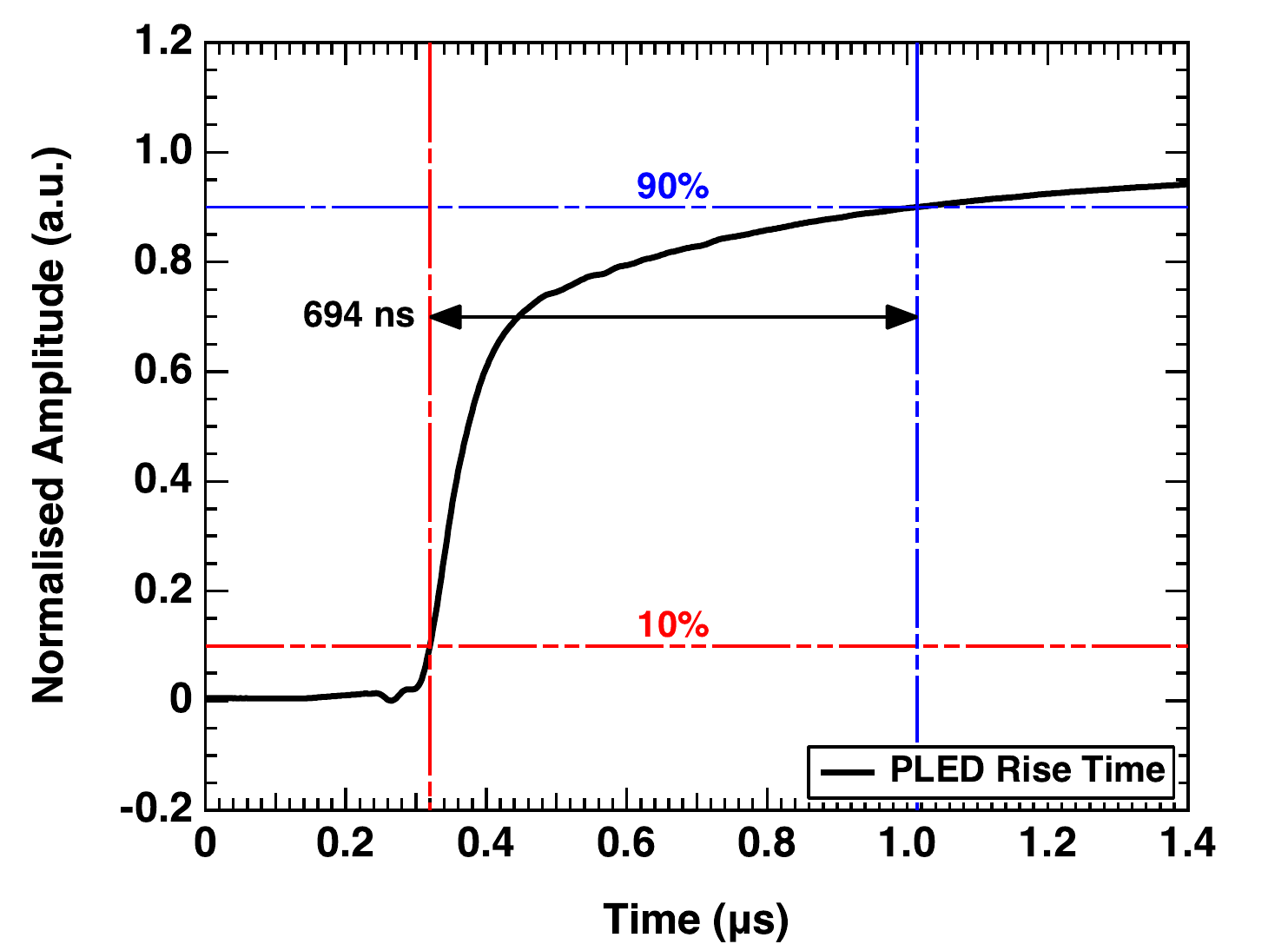}
  \caption{Measured rise time of the PLED, showing 694 ns}
  \label{fig:figure5}
\end{figure}

\begin{figure}[t]
  \centering
  \includegraphics[width=0.45\textwidth]{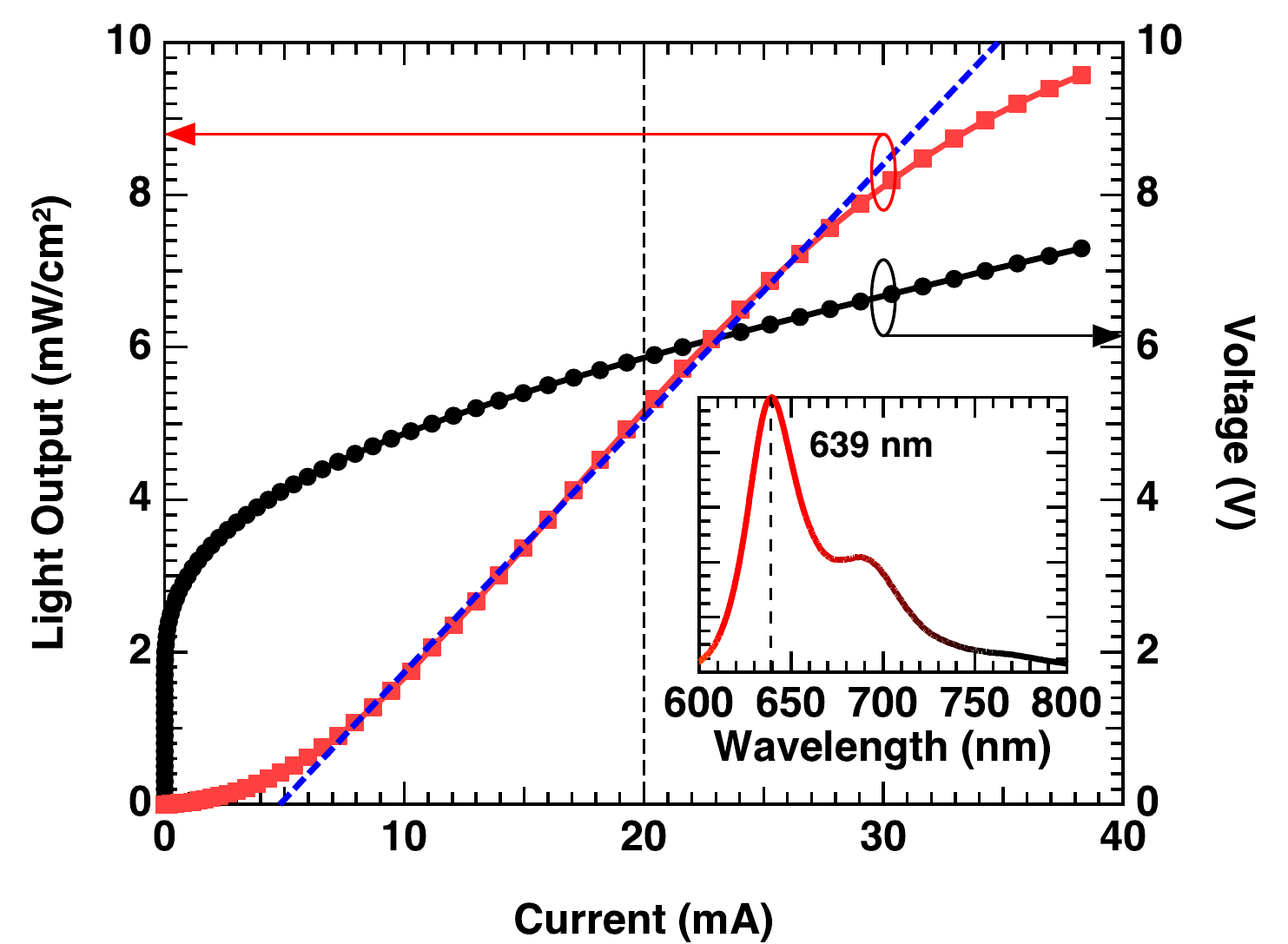}
  \caption{Measured L-I-V curve for the PLED under test and optical spectrum inset}
  \label{fig:figure6}
\end{figure}

The PLED bandwidth is estimated at $\sim$500~kHz based on the rise time measurement of 694~ns, see Fig.~\ref{fig:figure5}. This is in good agreement with the measured bandwidth values reported in our previous works \cite{haigh3,haigh2,le}. The measured PLED light output-current ($L$-$I$) and current-voltage ($I$-$V$) relationships are illustrated in Fig.~\ref{fig:figure6}, where the pseudo-linear region has been fitted with a linear curve. The PLED is biased at 20~mA\textsubscript{DC} and 20~mA\textsubscript{AC} is used for the signal. The emission spectrum of the PLED, referred to in \cite{minotto} and shown inset in Fig.~\ref{fig:figure6}, peaks at 639~nm.

The transmission distance $d$ is rather short at 0.05~m, which is consistent with reports in the literature, since we have used a single $\sim$4.5~mm\textsubscript{2} sized device with a relatively low luminance level in order to demonstrate the proof of concept \cite{chen,haigh3,haigh2,le}. Note, matrices of pixels can be used to increase luminance and $d$, which we will implement in our future work. However, we do not expect the performance of the link to vary when scaling up to multiple pixels. The mathematical channel model of the free space link can be referred to in \cite{lee}, neglecting the multi-path component due to the both low data rate (note, the bandwidth of a typical room is $>$10~MHz \cite{miramirkhani}) and the transmission distance.

The receiver is composed of a silicon photodetector (ThorLabs PDA10A-2) and transimpedance amplifier (150~MHz bandwidth, and a gain of $5\times10^3$~V/A). The output of the optical receiver is further amplified by a gain of 10 times and sampled using a real time oscilloscope (RTO, Tektronix MSO70804C). A minimum of $5\times10^7$ samples are acquired for each modulation scheme under test at a rate of at least 10 samples/symbol. The received samples are then imported into MATLAB for demodulation and further signal processing.

\subsection{Staggered-CAP Demodulation}
The OOK and 4-CAP demodulation processes are identical to those widely reported in the literature \cite{haigh3,haigh,olmedo,haigh2} and therefore not covered in detail here. For the sCAP system, the regenerated signal is replicated and applied to the time-reversed matched filters $f_0(-n)$, $f_1(-n)$, $f_2(-n)$ and $f_3(-n)$, see Fig.~\ref{fig:figure1}. Next, the $T/2$ symbol delay is used for three of the channels to ensure that the sampling point is ubiquitous for the entire system. The signals are then down-sampled to 1 sample/symbol at a rate of $n_{ss}$ prior to being applied to the common phase error (CPE) correction and de-mapping modules from the relevant constellations to generate an estimate of the transmitted data stream and for bit-by-bit BER testing.

\section{Results}
\begin{figure}[t]
  \centering
  \includegraphics[width=0.45\textwidth]{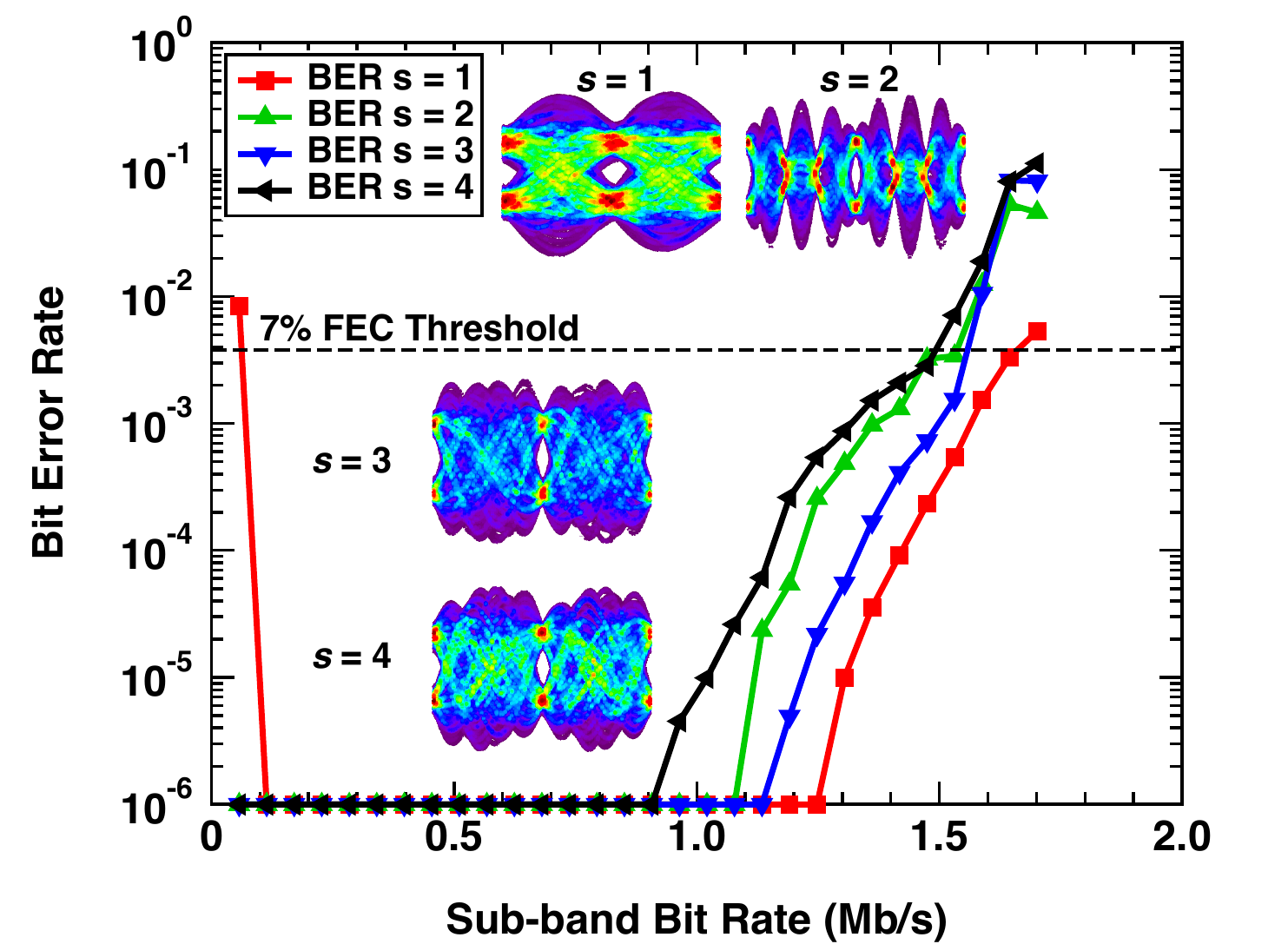}
  \caption{BER plots for each sub-band in the sCAP link, all eye diagrams shown are for a rate of 500~kb/s}
  \label{fig:figure7}
\end{figure}

The individual link BER performances as a function of the sub-band $s$ for sCAP are shown in Fig.~\ref{fig:figure7}. As shown, (\emph{i}) all sub-bands display error free performance below a sub-band data rate of 900~kb/s; and (\emph{ii}) at the 7\% forward error correction (FEC) of $3.8\times10^{-3}$ (shown as a dashed line in Fig.~\ref{fig:figure7}) the achievable sub-band rates are 1.7~Mb/s for $s=1$, 1.55~Mb/s for $s=2$ and $3$, and finally 1.48~Mb/s for $s=4$. The first sub-band, i.e. $s=1$ offers the best performance in terms of achievable un-coded data rate, at 1.25~Mb/s. This is because it suffers from the least out-of-band attenuation for of the any of the four sub-bands, since it occupies the lowest frequency range. Note, the higher BER ($10^{-2}$) measured at 50~kb/s is due to the baseline wander effect caused by a DC coupling capacitor. Sub-bands $s=2$ and $3$ are located in the same frequency range but separated by $\pi^c$ phase as in a generic CAP system. These two sub-bands are the next best performing, offering approximately 1.1 and 1.15~Mb/s for $s=2$ and $3$, respectively, i.e., approximately equivalent performance as expected. Clearly, as for $s=1$ they are less effected by the out-of-band attenuation than the final sub-band, which offers the worst performance at 0.9~Mb/s. The performance of $s=4$ is reliant on the matched filter, and hence the out-of-band attenuation distorts the signal shape, thus limiting the ability of the matched filter to remove the introduces inter-symbol interference (ISI).

\begin{figure}[t]
  \centering
  \includegraphics[width=0.45\textwidth]{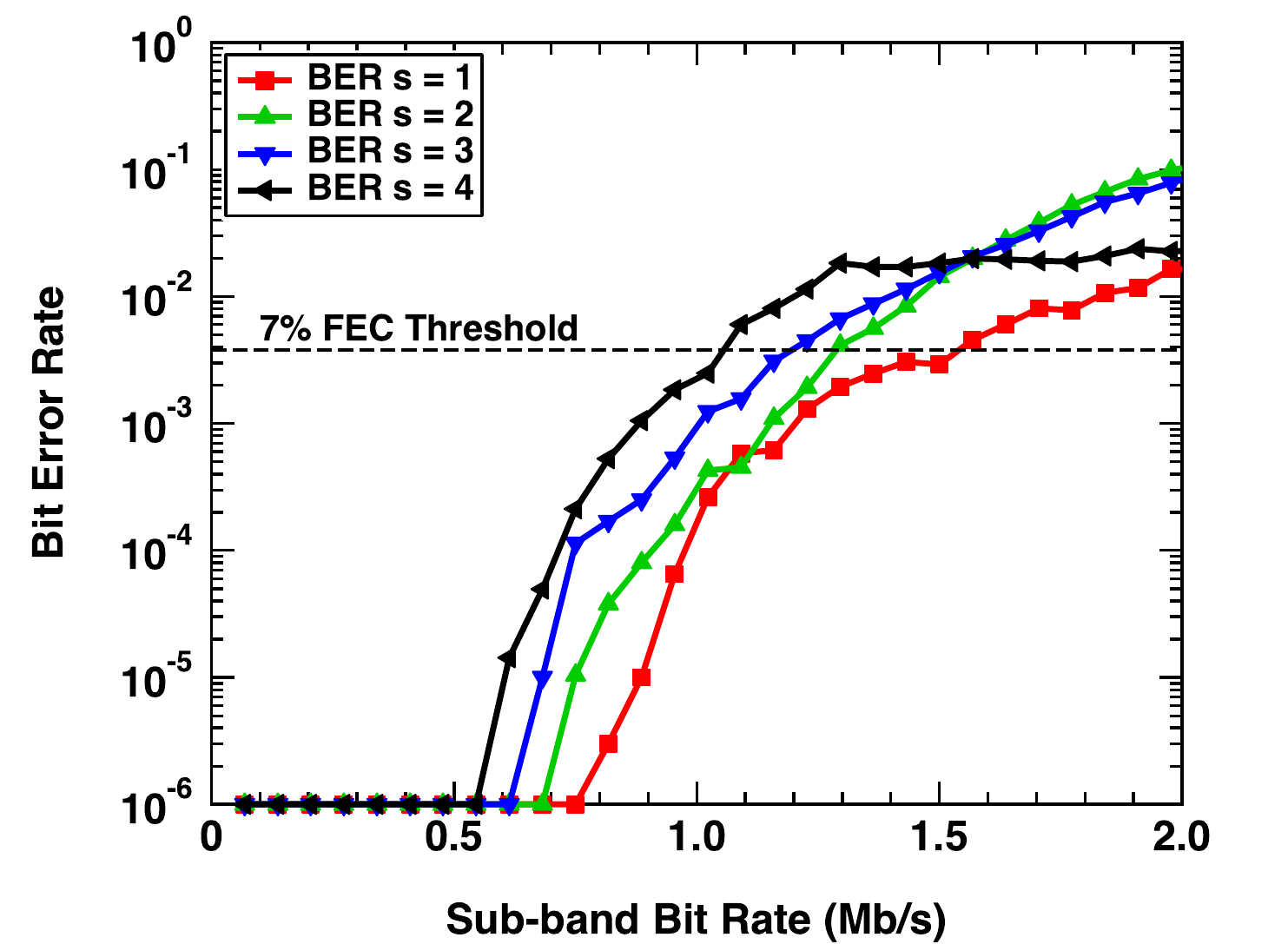}
  \caption{BER results for each sub-band in the 4-CAP link}
  \label{fig:figure8}
\end{figure}

An equivalent-energy conventional 4-CAP system was designed and transmitted in order to compare its performance with sCAP. Four sub-bands were selected to make an equivalent system as previously tested. The individual BER performance against the sub-band data rate is depicted in Fig.~\ref{fig:figure8}. It can clearly be seen that the BER performance of each sub-band is in the inverse order of sub-band frequency. This means that the lowest frequency sub-band (i.e., $s=1$) offers best performance because of reduced impact of out-of-band attenuation. An un-coded (FEC) data rate of 0.75 (1.5)~Mb/s can be supported. Sub-bands 2, 3 and 4 then follow, with un-coded (FEC) data rates of 0.65 (1.25)~Mb/s for $s=2$, 0.6 (1.1)~Mb/s for $s=3$ and 0.55 (1.05)~Mb/s for $s=4$.

\begin{figure}[t]
  \centering
  \includegraphics[width=0.45\textwidth]{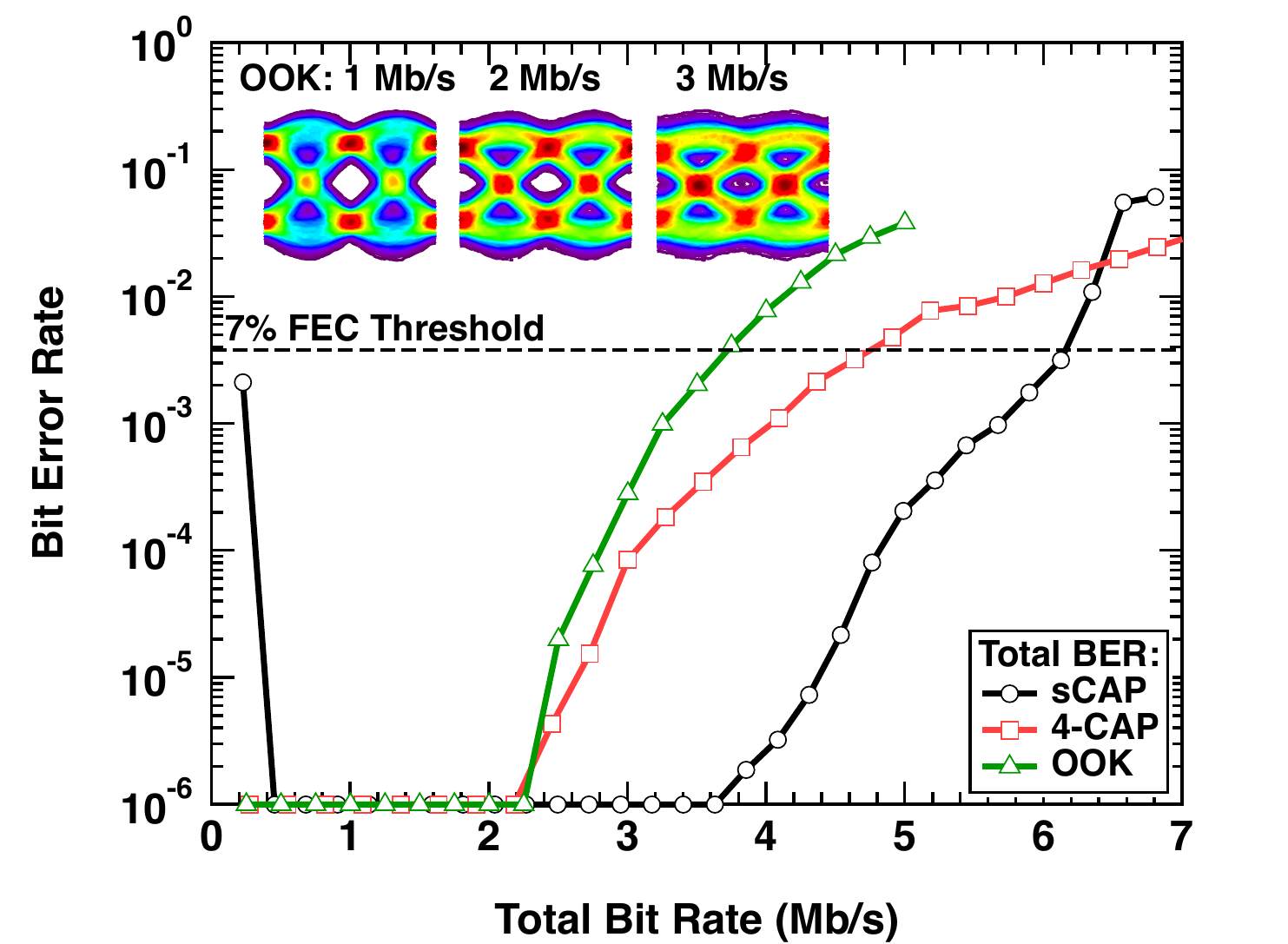}
  \caption{Comparative BER results for sCAP, 4-CAP and OOK}
  \label{fig:figure9}
\end{figure}

The total BER (i.e., ratio of the sum of all of errors in all sub-bands to the total number of bits transmitted) for sCAP, 4-CAP and OOK is illustrated in Fig.~\ref{fig:figure9}. Also shown are the eye diagrams for at 1, 2 and 3~Mb/s OOK with RRC pulse shaping and detection via matched filtering. Note, OOK and conventional 4-CAP can support an un-coded data rate of 2.25~Mb/s at a BER of $10{-6}$, thus no advantage is obtained in adopting the significantly more complex CAP scheme without optimisation. This result agrees well with studies reported in the literature, where typical un-coded links with pulse-based modulation offer data rates  of 2--3~Mb/s \cite{haigh3,haigh2,le} (note, the current record is 3~Mb/s \cite{le}). On the other hand, for the sCAP scheme proposed in \cite{stepniak} offers the un-coded rate is 3.75~Mb/s, i.e., a 25\% improvement compared to \cite{haigh3,haigh2,le}. Considering the 7\% FEC limit of $3.8\times10^{-3}$, see dashed line in Fig.~\ref{fig:figure9}, sCAP offers higher data rate of 6.2~Mb/s compared to (\emph{i}) 4.6~Mb/s and 3.75~Mb/s for 4-CAP and OOK, respectively; and (\emph{ii}) 4.8~Mb/s for OQAM-based OFDM \cite{chen}. Thus sCAP, which is a spectrally efficient modulation scheme with no guard intervals as in 4-CAP, offers 30\% higher data rates than previously reported record of 4.8 Mb/s.

\section{Conclusion}
In this work, we have demonstrated a new record un-coded transmission speed with no equalisation in a PLED based VLC link. We have confirmed that sCAP outperforms 4-CAP and pulse-shaped OOK systems over the same transmission link and equivalent energy. Further improvements can be made, since only binary signalling was utilised on each sCAP channel hence, a further improvement in the data rate can be achieved using the appropriate bit-loading algorithm. We note that the 30\% performance improvement in spectral efficiency compared to reports in the literature is the due to full spectrum utilisation capability of sCAP. Future work will combine sCAP with high performance equalisation schemes to further improve the achievable transmission rates.

\bibliographystyle{IEEEtran}
\bibliography{IEEE_ICC}

\begin{thebibliography}{10}
\providecommand{\url}[1]{#1}
\csname url@samestyle\endcsname
\providecommand{\newblock}{\relax}
\providecommand{\bibinfo}[2]{#2}
\providecommand{\BIBentrySTDinterwordspacing}{\spaceskip=0pt\relax}
\providecommand{\BIBentryALTinterwordstretchfactor}{4}
\providecommand{\BIBentryALTinterwordspacing}{\spaceskip=\fontdimen2\font plus
\BIBentryALTinterwordstretchfactor\fontdimen3\font minus
  \fontdimen4\font\relax}
\providecommand{\BIBforeignlanguage}[2]{{%
\expandafter\ifx\csname l@#1\endcsname\relax
\typeout{** WARNING: IEEEtran.bst: No hyphenation pattern has been}%
\typeout{** loaded for the language `#1'. Using the pattern for}%
\typeout{** the default language instead.}%
\else
\language=\csname l@#1\endcsname
\fi
#2}}
\providecommand{\BIBdecl}{\relax}
\BIBdecl

\bibitem{degans}
B.~De~Gans, P.~C. Duineveld, and U.~S. Schubert, ``Inkjet printing of polymers:
  state of the art and future developments,'' \emph{Adv Mater}, vol.~16, no.~3,
  pp. 203--213, 2004.

\bibitem{yimsiri}
P.~Yimsiri and M.~R. Mackley, ``Spin and dip coating of light-emitting polymer
  solutions: Matching experiment with modelling,'' \emph{Chemical Engineering
  Science}, vol.~61, no.~11, pp. 3496--3505, 2006.

\bibitem{gaynor}
W.~Gaynor, S.~Hofmann, M.~G. Christoforo, C.~Sachse, S.~Mehra, A.~Salleo, M.~D.
  McGehee, M.~C. Gather, B.~Lussem, L.~Muller-Meskamp, P.~Peumans, and K.~Leo,
  ``Color in the corners: {ITO}-free white {OLEDs} with angular color
  stability,'' \emph{Adv Mater}, vol.~25, no.~29, pp. 4006--13, 2013.

\bibitem{clayden}
J.~Clayden, N.~Greeves, and S.~Warren, \emph{Organic Chemistry}.\hskip 1em plus
  0.5em minus 0.4em\relax {OUP} Oxford, 2012.

\bibitem{clark}
J.~Clark and G.~Lanzani, ``Organic photonics for communications,'' \emph{Nature
  Photonics}, vol.~4, no.~7, pp. 438--446, 2010.

\bibitem{cisco}
, ``Global mobile data traffic forecast update, 2013-2018,'' \emph{Cisco, Cisco
  Visual Networking Index White Paper}, 2014.

\bibitem{wang2}
Y.~G. Wang, L.~Tao, X.~X. Huang, J.~Y. Shi, and N.~Chi, ``{8-Gb/s} {RGBY}
  {LED}-based {WDM} {VLC} system employing high-order {CAP} modulation and
  hybrid post equalizer,'' \emph{{IEEE} Photonics Journal}, vol.~7, no.~6, pp.
  1--7, 2015.

\bibitem{chun}
H.~Chun, S.~Rajbhandari, G.~Faulkner, D.~Tsonev, E.~Y. Xie, J.~J.~D. McKendry,
  E.~D. Gu, M.~D. Dawson, D.~C. O'Brien, and H.~Haas, ``{LED} based wavelength
  division multiplexed {10 Gb/s} visible light communications,'' \emph{Journal
  of Lightwave Technology}, vol.~34, no.~13, pp. 3047--3052, 2016.

\bibitem{islim}
M.~S. Islim, R.~X. Ferreira, X.~Y. He, E.~Y. Xie, S.~Videv, S.~Viola,
  S.~Watson, N.~Bamiedakis, R.~V. Penty, I.~H. White, A.~E. Kelly, E.~D. Gu,
  H.~Haas, and M.~D. Dawson, ``Towards {10 Gb/s} orthogonal frequency division
  multiplexing-based visible light communication using a {GaN} violet
  {micro-LED},'' \emph{Photonics Research}, vol.~5, no.~2, pp. A35--A43, 2017.

\bibitem{wang}
Y.~Wang, X.~Huang, L.~Tao, J.~Shi, and N.~Chi, ``{4.5-Gb/s} {RGB-LED} based
  {WDM} visible light communication system employing {CAP} modulation and {RLS}
  based adaptive equalization,'' \emph{Opt Express}, vol.~23, no.~10, pp.
  13\,626--33, 2015.

\bibitem{das}
R.~Das, K.~Ghaffarzadeh, and X.~He, ``Global {OLED} display forecasts and
  technologies 2019-2029: The rise of flexible and foldable displays,''
  \emph{IDTechEx}, 2018.

\bibitem{boccardi}
F.~Boccardi, J.~Andrews, H.~Elshaer, M.~Dohler, S.~Parkvall, P.~Popovski, and
  S.~Singh, ``Why to decouple the uplink and downlink in cellular networks and
  how to do it,'' \emph{{IEEE} Communications Magazine}, vol.~54, no.~3, pp.
  110--117, 2016.

\bibitem{shao}
S.~H. Shao, A.~Khreishah, M.~Ayyash, M.~B. Rahaim, H.~Elgala, V.~Jungnickel,
  D.~Schulz, T.~D.~C. Little, J.~Hilt, and R.~Freund, ``Design and analysis of
  a visible-light-communication enhanced {WiFi} system,'' \emph{Journal of
  Optical Communications and Networking}, vol.~7, no.~10, pp. 960--973, 2015.

\bibitem{shao2}
S.~Shao, A.~Khreishah, M.~B. Rahaim, H.~Elgala, M.~Ayyash, T.~D.~C. Little, and
  J.~Wu, ``An indoor hybrid {WiFi-VLC} internet access system,'' in \emph{2014
  {IEEE} 11th International Conference on Mobile Ad Hoc and Sensor Systems},
  Conference Proceedings, pp. 569--574.

\bibitem{chen}
H.~J. Chen, Z.~Y. Xu, Q.~Gao, and S.~B. Li, ``A {51.6 Mb/s} experimental {VLC}
  system using a monochromic organic {LED},'' \emph{{IEEE} Photonics Journal},
  vol.~10, no.~2, pp. 1--12, 2018.

\bibitem{haigh3}
P.~A. Haigh, F.~Bausi, H.~Le~Minh, I.~Papakonstantinou, W.~O. Popoola,
  A.~Burton, and F.~Cacialli, ``Wavelength-multiplexed polymer {LEDs}: Towards
  {55 Mb/s} organic visible light communications,'' \emph{{IEEE} Journal on
  Selected Areas in Communications}, vol.~33, no.~9, pp. 1819--1828, 2015.

\bibitem{gurram}
S.~Gurram, O.~Brennan, and T.~Wilkerson, ``{DC-to-DC} switching-regulator
  insights–achieving longer battery life in {DSP} systems,'' \emph{Analog
  Dialogue}, vol.~41, no.~4, pp. 11--15, 2007.

\bibitem{wu}
F.~M. Wu, C.~T. Lin, C.~C. Wei, C.~W. Chen, Z.~Y. Chen, H.~T. Huang, and
  S.~Chi, ``Performance comparison of {OFDM} signal and {CAP} signal over high
  capacity {RGB-LED-Based WDM} visible light communication,'' \emph{{IEEE}
  Photonics Journal}, vol.~5, no.~4, pp. 7\,901\,507--7\,901\,507, 2013.

\bibitem{haigh}
P.~A. Haigh, A.~Burton, K.~Werfli, H.~L. Minh, E.~Bentley, P.~Chvojka, W.~O.
  Popoola, I.~Papakonstantinou, and S.~Zvanovec, ``A multi-cap visible-light
  communications system with {4.85-b/s/Hz} spectral efficiency,'' \emph{{IEEE}
  Journal on Selected Areas in Communications}, vol.~33, no.~9, pp. 1771--1779,
  2015.

\bibitem{stepniak}
G.~Stepniak, ``Staggered {CAP}-a new spectrally efficient modulation format for
  optical communications,'' \emph{{IEEE} Photonics Technology Letters},
  vol.~30, no.~4, pp. 367--370, 2018.

\bibitem{zhao}
J.~Zhao and A.~D. Ellis, ``Offset-{QAM} based coherent wdm for spectral
  efficiency enhancement,'' \emph{Optics Express}, vol.~19, no.~15, pp.
  14\,617--14\,631, 2011.

\bibitem{olmedo}
M.~I. Olmedo, T.~J. Zuo, J.~B. Jensen, Q.~W. Zhong, X.~G. Xu, S.~Popov, and
  I.~T. Monroy, ``Multiband carrierless amplitude phase modulation for high
  capacity optical data links,'' \emph{Journal of Lightwave Technology},
  vol.~32, no.~4, pp. 798--804, 2014.

\bibitem{haigh2}
P.~A. Haigh, F.~Bausi, Z.~Ghassemlooy, I.~Papakonstantinou, H.~Le~Minh,
  C.~Flechon, and F.~Cacialli, ``Visible light communications: real time {10
  Mb/s} link with a low bandwidth polymer light-emitting diode,'' \emph{Opt
  Express}, vol.~22, no.~3, pp. 2830--8, 2014.

\bibitem{slim}
I.~Slim, A.~Mezghani, L.~G. Baltar, J.~Qi, and J.~A. Nossek, ``Frequency domain
  vs. time domain filter design of {RRC} pulse shaper for spectral confinement
  in high speed optical communications,'' in \emph{2013 {ITG} Symposium
  Proceedings - Photonic Networks}, Conference Proceedings, pp. 1--3.

\bibitem{joost}
M.~Joost, ``Theory of root-raised cosine filter,'' \emph{Research and
  Development}, vol. 47829, 2010.

\bibitem{brown}
T.~M. Brown, G.~M. Lazzerini, L.~J. Parrott, V.~Bodrozic, L.~Bürgi, and
  F.~Cacialli, ``Time dependence and freezing-in of the electrode oxygen
  plasma-induced work function enhancement in polymer semiconductor
  heterostructures,'' \emph{Organic Electronics}, vol.~12, no.~4, pp. 623--633,
  2011.

\bibitem{minotto}
A.~Minotto, P.~Murto, Z.~Genene, A.~Zampetti, G.~Carnicella, W.~Mammo, M.~R.
  Andersson, E.~Wang, and F.~Cacialli, ``Efficient near-infrared
  electroluminescence at 840 nm with {"Metal-Free"} {Small-Molecule:Polymer
  Blends},'' \emph{Adv Mater}, vol.~30, no.~34, p. e1706584, 2018.

\bibitem{le}
S.~T. Le, T.~Kanesan, F.~Bausi, P.~A. Haigh, S.~Rajbhandari, Z.~Ghassemlooy,
  I.~Papakonstantinou, W.~O. Popoola, A.~Burton, H.~Le~Minh, F.~Cacialli, and
  A.~D. Ellis, ``{10 Mb/s} visible light transmission system using a polymer
  light-emitting diode with orthogonal frequency division multiplexing,''
  \emph{Optics Letters}, vol.~39, no.~13, pp. 3876--3879, 2014.

\bibitem{lee}
K.~Lee, H.~Park, and J.~R. Barry, ``Indoor channel characteristics for visible
  light communications,'' \emph{{IEEE} Communications Letters}, vol.~15, no.~2,
  pp. 217--219, 2011.

\bibitem{miramirkhani}
F.~Miramirkhani and M.~Uysal, ``Channel modeling and characterization for
  visible light communications,'' \emph{{IEEE} Photonics Journal}, vol.~7,
  no.~6, pp. 1--16, 2015.

\end{thebibliography}
\bstctlcite{IEEE_ICC}

\end{document}